\documentclass[noshowpacs,twocolumn,prd,amsfonts,amssymb]{revtex4}

\begin{document}

\title{Remarks on the conformal transformations}
\author{L. C. T. Guillen}
\email{lctorres@ift.unesp.br}

\affiliation{Instituto de F\'{\i}sica Te\'orica \\
Universidade Estadual Paulista \\
Rua Pamplona 145 \\ 01405-900 S\~ao Paulo SP, Brazil}

\date{\today}

\begin{abstract}
Conformal transformations are obtained by demanding that the form of the 
metric change by a ``conformal'' factor. Nevertheless, in the literature, 
this transformation of the metric is not taken into account when a 
variation of the action is performed. As a consequence, it is obtained 
that massive particles are not invariant under the conformal 
transformations, and that the scale dimension $d$ of the fields 
coincides with the natural dimension of the fields, in the sense of 
dimensional analysis. The basic purpose of this paper is to take the 
transformation of the metric into the variation of the action. When this 
is done, we obtain now that even  massive particles are invariant under the 
conformal transformations.  Also, the scale dimension $d$ of the fields does 
not coincides anymore with the natural dimension of the fields, but seems 
to be related with the tensorial character of the fields.

\end{abstract}

\maketitle

\section{Introduction}

Besides invariant under the Poincar\'e transformations (translations and 
Lorentz transformations), Maxwell's electromagnetic action possesses a 
wider invariance, namely, invariance under the conformal 
transformations. They were first obtained by Cunningham and 
Bateman~\cite{cuba}. (For a historical review see~\cite{frw}.) Since 
then, the conformal transformations have been widely studied. 
(See~\cite{wess,masa,cacoja,wess2,carru,coja,fgg} and references there 
in.) In all these works it is asserted that the masses of particles 
breaks dilatation, and consequently also conformal, invariance. It is 
a well known result too, that quantically, dilatation can not be a 
symmetry of nature, because it would imply that all masses vanishes, 
or that the mass spectrum is continuous~\cite{wess}. 

Nevertheless, although the conformal transformations are just obtained 
imposing that the form of the metric change by a ``conformal'' factor, it seems 
strange that it is usually assumed that the metric does not change. 
In fact, in~\cite{wess2,blago} they give and argument to 
suppose that. In this paper we will not use this argument because it 
seems to violate the isometries of the metric. As is well known, an 
isometry is a coordinate transformation which leaves the form of the 
metric invariant~\cite{weinberg}. For the case of the Minkowski metric, 
only the Poincar\'e transformations are the isometries of the metric 
(Killing vectors), while the conformal transformations are not.

Consequently, in this paper we will take the change of the metric 
seriously, and re-examine the question of the conformal 
transformations. For example, in the variation of the Lagrangian, besides 
the usual variation of the fields, we have to take also the variation of 
the metric. When the transformation is an isometry of the metric, this 
variation vanishes. But, for transformations that are not isometries, 
the variation of the metric introduces one extra term, which is 
proportional to the symmetric energy-momentum tensor. This extra term 
implies that Noether theorem no longer gives a conserved current even 
when the action is invariant under this transformation. Only when the 
trace of the symmetric energy-momentum tensor vanishes we obtain a 
conserved current. Also, massive particles turn out now to be invariant 
under the conformal transformations. This is because for massive 
particles the trace of the symmetric energy-momentum tensor has a mass 
term which cancels the usual mass term that would appear. The invariance 
condition of the action under the conformal transformations, implies 
that the scale dimension $d$ is related with the tensorial character of 
the fields.

We organize the paper as follows: in Sec. II we give just a brief review 
of the conformal transformations. Then, in Sec. III we introduce the 
variation of the metric and re-obtain Noether theorem for this case. 
Based on this, we reconsider the conformal transformations in Sec. IV. We 
apply these new considerations to the massive scalar and spinor field, 
and to the electromagnetic field in Sec. V. Finally, we draw the main 
conclusions of the paper in Sec. VI.

\section{Conformal Transformations}

We will use the Greek alphabet ($\mu, \nu, \rho, \dots = 0, \dots, 3$) 
to denote spacetime indices. In the following, we are going to work in 
the usual four-dimensional Minkowski spacetime, with the Minkowski 
metric $\eta_{\mu \nu} = \mbox{diag}~(1, - 1, - 1, - 1)$.

The conformal transformations are obtained by demanding that the metric 
changes its form by
\begin{equation}
g_{\mu \nu} (x) = \lambda (x) \eta_{\mu \nu} .
\end{equation}
We easily see that this transformation of the metric leave the 
light-cone $d s^2 = \eta_{\mu \nu} d x^{\mu} d x^{\nu} = 0$ invariant. 
Infinitesimally, $\lambda (x) = 1 + \Omega (x)$, the change in the 
form of the metric is
\begin{equation}
\bar{\delta} \eta_{\mu \nu} = g_{\mu \nu} (x) - \eta_{\mu \nu} = 
\Omega (x) \eta_{\mu \nu} .   \label{contrame}
\end{equation}

Now, under the infinitesimal coordinate transformation
\begin{equation}
x^{\prime \mu} = x^{\mu} + \delta x^{\mu} (x) ,   \label{ct}
\end{equation}
Minkowski metric $\eta_{\mu \nu}$ changes its form according to
\begin{equation}
\bar{\delta} \eta_{\mu \nu} = g_{\mu \nu} (x) - \eta_{\mu \nu} = - 
\eta_{\mu \rho} \partial_{\nu} \delta x^{\rho} - \eta_{\nu \rho} 
\partial_{\mu} \delta x^{\rho} .   \label{tramime}
\end{equation}
We should remark that although Minkowski metric $\eta_{\mu \nu}$ is 
constant, the coordinate transformed metric $g_{\mu \nu}$ is not 
necessarilly equal to $\eta_{\mu \nu}$, nor even constant. Under an 
arbitrary coordinate transformation, the metric $g_{\mu \nu}$ will in 
general depend on the spacetime coordinates. Therefore, when 
we talk about the variation of the Minkowski metric, we mean according to 
(\ref{tramime}) the difference at the same point between the coordinate 
transformed metric $g_{\mu \nu}$ and Minkowski metric $\eta_{\mu \nu}$. 
Then, substituting (\ref{contrame}) in (\ref{tramime}), we 
arrive at the following equation,
\begin{equation}
\eta_{\mu \rho} \partial_{\nu} \delta x^{\rho} + \eta_{\nu \rho} 
\partial_{\mu} \delta x^{\rho} = \frac{1}{2} \eta_{\mu \nu} 
\partial_{\rho} \delta x^{\rho} ,   \label{cke}
\end{equation}
where $\Omega (x) = - (\frac{1}{2}) \partial_{\mu} \delta x^{\mu}$. 
This is called the {\it conformal Killing equation}. The Poincar\'e 
transformations are a particular solution to this equation with 
$\Omega (x) = 0$, that is, they are solution of the Killing equation 
$\bar{\delta} \eta_{\mu \nu} = 0$. 
The solutions of the conformal Killing equation (\ref{cke}) are the 
conformal transformations
\begin{equation}
\delta_D x^{\mu} = a x^{\mu} ,   \label{dila}
\end{equation}
which are called the {\it dilatations}, and
\begin{equation}
\delta_S x^{\mu} = 2 x^{\mu} c_{\nu} x^{\nu} - c^{\mu} x_{\nu} 
x^{\nu} ,   \label{sct}
\end{equation}
which are called the {\it special conformal transformations}. (For a 
good review on the conformal transformations see~\cite{blago}.) 
Therefore, the conformal transformations are coordinate transformations, 
as can be seen from the {\it l.h.s.} of (\ref{cke}), which changes the 
form of the metric according to (\ref{contrame}), the {\it r.h.s.} of 
(\ref{cke}).

\section{The Variation of the Metric and Noether Theorem}

Now, comes two important and crucial questions: Is Minkowski metric 
$\eta_{\mu \nu}$ invariant under the conformal transformations? 
Secondly, if Minkowski metric $\eta_{\mu \nu}$ is not invariant under 
the conformal transformations, Should we take the variation of the 
metric when we vary the action?

Usually in the 
literature~\cite{wess,masa,cacoja,wess2,carru,coja,fgg,blago}, it 
is assumed that Minkowski metric $\eta_{\mu \nu}$ does not changes 
under the conformal transformations. In fact, in~\cite{wess2,blago} 
they use the Weyl rescaling together with the conformal 
transformations to impose that $\bar{\delta} \eta_{\mu \nu} = 0$. As 
remarked in~\cite{wess2}, without the Weyl rescaling we can not obtain that 
$\bar{\delta} \eta_{\mu \nu} = 0$ under the conformal transformations. 
(Note that as we are considering coordinate transformations, by 
$\bar{\delta} \eta_{\mu \nu}$ we mean the difference between $g_{\mu \nu}$ 
and $\eta_{\mu \nu}$, as explained below (\ref{tramime}), and this 
difference will in general not vanishes.)
Besides that, it seems strange that the Weyl rescaling is used just to 
the metric, and not to the other fields too. In this work we will not 
use the Weyl rescaling argument, and therefore Minkowski metric 
$\eta_{\mu \nu}$ is not invariant under the conformal transformations. 
This seems to be the natural answer, because the conformal 
transformations (\ref{dila}) and (\ref{sct}) {\it satisfy} the conformal 
Killing equation (\ref{cke}), and not a killing equation. Only 
translations and Lorentz transformations are the Killing vectors of 
$\eta_{\mu \nu}$~\cite{weinberg}. For example, substituting  (\ref{dila}) 
in (\ref{tramime}) we obtain that
\begin{equation}
\bar{\delta}_D \eta_{\mu \nu} = - 2 a \eta_{\mu \nu} ,  \label{divame}
\end{equation}
which is of the form (\ref{contrame}) with $\Omega (x) = - 2 a$. Now, 
substituting (\ref{sct}) in (\ref{tramime}) we obtain that
\begin{equation}
\bar{\delta}_S \eta_{\mu \nu} = - 4 c_{\rho} x^{\rho} \eta_{\mu \nu} ,
\label{sctvame}
\end{equation}
which is of the form (\ref{contrame}) with 
$\Omega (x) = - 4 c_{\rho} x^{\rho}$. Although Minkowski's metric is 
not a dynamical field, it changes under the conformal transformations, 
and we should take this change of the metric when we make a variation 
in the action.

Let us begin, then, giving the action of a general field $\Phi$,
\begin{equation}
{\mathcal A} = \int d^4 x \; {\mathcal L} (\Phi, \partial_{\mu} \Phi),
\label{action}
\end{equation}
where ${\mathcal L} (\Phi, \partial_{\mu} \Phi)$ is the Lagrangian of 
the field $\Phi$, and it depends only on the field and its first 
derivative. Under the infinitesimal coordinate transformations 
(\ref{ct}), we have a transformation in the field $\Phi$ given 
by~\cite{blago,ramond}
\begin{equation}
\delta \Phi (x) = \Phi^{\prime} (x^{\prime}) - \Phi (x) = \bar{\delta} 
\Phi (x) + \delta x^{\mu} \partial_{\mu} \Phi (x) ,   \label{tvf}
\end{equation}
where $\bar{\delta} \Phi (x) = \Phi^{\prime} (x) - \Phi (x)$ is a 
variation just in the form of the field. The transformations 
(\ref{ct}) and (\ref{tvf}) induces the following transformation on the 
Lagrangian,
\begin{equation}
\delta {\mathcal L} = \bar{\delta} {\mathcal L} + 
\delta x^{\mu} \partial_{\mu} {\mathcal L} ,   \label{fpvl}
\end{equation}
where
\begin{equation}
\bar{\delta} {\mathcal L} = 
\frac{\partial {\mathcal L}}{\partial \Phi} \bar{\delta} \Phi + 
\frac{\partial {\mathcal L}}{\partial (\partial_{\mu} \Phi)} 
\bar{\delta} (\partial_{\mu} \Phi) + 
\frac{\partial {\mathcal L}}{\partial \eta_{\mu \nu}} \bar{\delta} 
\eta_{\mu \nu} ,   \label{newfvl}
\end{equation}
is the variation in the form of the Lagrangian. Note that we have also 
taken the variation in the form of the metric. This term just vanishes 
for isometries transformations. As is well known, the variation of 
the Lagrangian with respect to the metric is
\begin{equation}
\frac{\partial {\mathcal L}}{\partial \eta_{\mu \nu}} = - \frac{1}{2} 
\sqrt{- \eta} \, {\mathcal T}^{\mu \nu} ,   \label{vlsemt}
\end{equation}
where $\eta = \det (\eta_{\mu \nu})$, and ${\mathcal T}^{\mu \nu}$ is 
the {\it Symmetric Energy-Momentum Tensor (SEMT)}, that is, the 
{\it Canonical Energy-Momentum Tensor (CEMT)} symmetrized through the 
Belinfante procedure~\cite{belinfante},
\begin{equation}
{\mathcal T}^{\mu \nu} = t^{\mu \nu} - \frac{1}{2} \partial_{\rho} 
\varphi^{\rho \mu \nu} ,   \label{semt}
\end{equation}
where
\begin{equation}
t^{\mu}{}_{\nu} = 
\frac{\partial {\mathcal L}}{\partial (\partial_{\mu} \Phi)} 
\partial_{\nu} \Phi - \delta^{\mu}{}_{\nu} {\mathcal L} , \label{cemt}
\end{equation}
is the definition of the {\it CEMT},
\begin{equation}
\varphi^{\rho \mu \nu} = - \varphi^{\mu \rho \nu} = 
{\mathcal S}^{\rho \mu \nu} + {\mathcal S}^{\mu \nu \rho} - 
{\mathcal S}^{\nu \rho \mu} ,    \label{sss}
\end{equation}
with
\begin{equation}
{\mathcal S}^{\rho}{}_{\mu \nu} = \mathrm{i} 
\frac{\partial {\mathcal L}}{\partial (\partial_{\rho} \Phi)} 
S_{\mu \nu} \Phi ,   \label{spintgf}
\end{equation}
the definition of the {\it Spin} Tensor, and $S_{\mu \nu}$ the spin 
generator in an appropriate representation to the field 
$\Phi$~\cite{masa,blago,ramond}.

As $\delta d^4 x = \partial_{\mu} \delta x^{\mu} d^4 x$, and using 
(\ref{fpvl}), the invariance of the action under the transformations 
(\ref{ct}) and (\ref{tvf}) implies that~\cite{blago}
\begin{equation}
\Delta {\mathcal L} = {\mathcal L} \partial_{\mu} \delta 
x^{\mu} + \bar{\delta} {\mathcal L} + \delta x^{\mu} 
\partial_{\mu} {\mathcal L} = 0 .   \label{newtvl}
\end{equation}
This is a condition involving the Lagrangian which must be satisfied 
if we want $\delta {\mathcal A} = 0$ for some symmetry transformation. 
As $\bar{\delta} (\partial_{\mu} \Phi) = \partial_{\mu} (\bar{\delta} 
\Phi)$, then, doing a partial integration in (\ref{newfvl}) and 
substituting in (\ref{newtvl}), we obtain a similar form of the usual 
Noether theorem,
\begin{equation}
\frac{\delta {\mathcal L}}{\delta \Phi} \bar{\delta} \Phi - 
\partial_{\mu} J^{\mu} = - 
\frac{\partial {\mathcal L}}{\partial \eta_{\mu \nu}} \bar{\delta} 
\eta_{\mu \nu} ,   \label{ncc}
\end{equation}
where
\begin{equation}
\frac{\delta {\mathcal L}}{\delta \Phi} = 
\frac{\partial {\mathcal L}}{\partial \Phi} - \partial_{\mu} 
\left(\frac{\partial {\mathcal L}}{\partial (\partial_{\mu} \Phi)} 
\right) ,  \label{elfunvar}
\end{equation}
is the Euler-Lagrange functional variation, and $J^{\mu}$ is the 
definition of the current,
\begin{equation}
J^{\mu} = - 
\frac{\partial {\mathcal L}}{\partial (\partial_{\mu} \Phi)} 
\bar{\delta} \Phi - {\mathcal L} \delta x^{\mu} .   \label{cc}
\end{equation}
When the transformation is an isometry of the metric, we easily 
see that the {\it r.h.s.} in (\ref{ncc}) vanishes. Then, provided 
that we are in the field equations, 
$\delta {\mathcal L} / \delta \Phi = 0$, the current $J^{\mu}$ is 
conserved, $\partial_{\mu} J^{\mu} = 0$. Therefore, invariance of the 
action under some isometry transformation implies a conserved 
current~\cite{kopov}. For example, invariance of the action under 
translations implies that the conserved current is the {\it CEMT} 
(\ref{cemt}), while invariance of the action under Lorentz 
transformations implies that the conserved current is the 
{\it Total Angular Momentum Tensor}
\begin{equation}
{\mathcal J}^{\mu}{}_{\rho \sigma} = L^{\mu}{}_{\rho \sigma} + 
{\mathcal S}^{\mu}{}_{\rho \sigma} ,   \label{tamt}
\end{equation}
where $L^{\mu}{}_{\rho \sigma} = x_{\sigma} t^{\mu}{}_{\rho} - 
x_{\rho} t^{\mu}{}_{\sigma}$ is the {\it Orbital Angular Momentum 
Tensor}. But, for transformations that are not isometries of the 
metric, there is the extra term in the {\it r.h.s.} of (\ref{ncc}). 
Therefore, in this case, we can see from (\ref{ncc}) that even if the 
transformation is a symmetry of the action, $\delta {\mathcal A} = 0$, 
and when we are in the field equations, 
$\delta {\mathcal L} / \delta \Phi = 0$, this does not necessarily 
means that we have a conserved current. As the conformal 
transformations are of this kind, let us see how the variation of the 
metric modifies the earlier considerations on conformal 
transformations.

\section{Conformal Transformations and Noether Theorem}

\subsection{Dilatations}

Under the dilatations transformations (\ref{dila}), the general 
transformation law of the field $\Phi$ 
is~\cite{masa,cacoja,carru,coja,fgg,blago}
\begin{equation}
\bar{\delta}_D \Phi = - a (x^{\mu} \partial_{\mu} \Phi + d \Phi) ,
\label{fvpd}
\end{equation}
where $d$ is the {\it scale dimension} of the field $\Phi$. We can 
write (\ref{divame}) in the form of (\ref{fvpd}) provided that 
$d (\eta_{\mu \nu}) = 2$. Note that as remarked in~\cite{carru,coja}, 
we should not confuse the scale dimension of the field with its natural 
dimension, in the sense of dimensional analysis. Therefore, the fact 
that we are choosing $d (\eta_{\mu \nu}) = 2$, does not mean that we are 
assigning a natural dimension to the metric. Then, substituting 
(\ref{fvpd}) in (\ref{newfvl}) and using (\ref{vlsemt}) and 
(\ref{divame}), we obtain that
\begin{equation}
\bar{\delta}_D {\mathcal L} = - a x^{\mu} \partial_{\mu} 
{\mathcal L} - a d \frac{\partial {\mathcal L}}{\partial \Phi} \Phi - 
a (d + 1) \frac{\partial {\mathcal L}}{\partial (\partial_{\mu} \Phi)} 
\partial_{\mu} \Phi + a {\mathcal T}^{\mu}{}_{\mu} ,   \label{newfvl2}
\end{equation}
where ${\mathcal T}^{\mu}{}_{\mu}$ is just the trace of the 
{\it SEMT}, and where we used that the action is invariant under 
translations~\cite{cacoja,coja,blago},
\begin{equation}
\partial_{\mu} {\mathcal L} = 
\frac{\partial {\mathcal L}}{\partial \Phi} \partial_{\mu} \Phi + 
\frac{\partial {\mathcal L}}{\partial (\partial_{\nu} \Phi)} 
\partial_{\mu} \partial_{\nu} \Phi .   \label{pdl}
\end{equation}
Then, substituting (\ref{newfvl2}) in (\ref{newtvl}) and using 
(\ref{dila}), the variation of the action is
\begin{equation}
\Delta_D {\mathcal L} = 4 a {\mathcal L} - a d 
\frac{\partial {\mathcal L}}{\partial \Phi} \Phi - a (d + 1) 
\frac{\partial {\mathcal L}}{\partial (\partial_{\mu} \Phi)} 
\partial_{\mu} \Phi + a {\mathcal T}^{\mu}{}_{\mu} .  \label{newtvld}
\end{equation}
Invariance of the action under dilatations, 
$\Delta_D {\mathcal L} = 0$, implies that
\begin{equation}
4 {\mathcal L} = d \frac{\partial {\mathcal L}}{\partial \Phi} \Phi + 
(d + 1) \frac{\partial {\mathcal L}}{\partial (\partial_{\mu} \Phi)} 
\partial_{\mu} \Phi - {\mathcal T}^{\mu}{}_{\mu} .   \label{newpdl3}
\end{equation}
Without the last term, this equation is usually interpreted as saying 
that only massless fields are invariant under 
dilatations~\cite{masa,coja,fgg}. But, as we will see below, the last term 
just cancels the mass terms, and we can say that dilatation invariance 
tell us how we can ``decompose'' the Lagrangian. Now, substituting 
(\ref{newpdl3}) in (\ref{newfvl2}), we obtain
\begin{equation}
\bar{\delta}_D {\mathcal L} = - a x^{\mu} \partial_{\mu} 
{\mathcal L} - 4 a {\mathcal L} .   \label{newfvl3}
\end{equation}
We see that if the action is invariant under dilatations, the 
variation of the Lagrangian can be written as in (\ref{fvpd}) with 
$d (\mathcal L) = 4$. Now, comes an important point. The number $4$ 
appearing in the above variation came from the term ${\mathcal L} 
\partial_{\mu} \delta x^{\mu} = {\mathcal L} \partial_{\mu} 
(a x^{\mu}) = 4 a {\mathcal L}$. Consequently, the number $4$ that 
appears in the variation of the Lagrangian does not seems to be related 
with the natural dimension of the Lagrangian. The number $4$ came from 
the dimension of the spacetime, which in this case is the 
four-dimensional Minkowski spacetime. Therefore, as we will see below, 
the parameter $d$ has nothing to do with the natural dimension of the 
fields when we consider the variation of the metric.

Now, substituting (\ref{dila}), (\ref{fvpd}), (\ref{divame}) and 
(\ref{vlsemt}) in (\ref{ncc}), and assuming that we are in the field 
equations, we obtain that
\begin{equation}
\partial_{\mu} D^{\mu} = {\mathcal T}^{\mu}{}_{\mu} ,  \label{codicu}
\end{equation}
where
\begin{equation}
D^{\mu} = x^{\nu} t^{\mu}{}_{\nu} + d 
\frac{\partial {\mathcal L}}{\partial (\partial_{\mu} \Phi)} \Phi ,
\label{dilacu}
\end{equation}
is the {\it dilatation current}. Therefore, the dilatation current is 
not conserved unless the trace of the {\it SEMT} vanishes. As the 
trace of the {\it SEMT} is proportional to the mass of the field, we 
see, in this way, why massive fields do not have a conserved 
dilatation current.

\subsection{Special conformal transformations}

The general transformation law of the field $\Phi$ under the special 
conformal transformations (\ref{sct}) 
is~\cite{masa,cacoja,carru,coja,fgg,blago}
\begin{equation}
\bar{\delta}_S \Phi = - c^{\mu} (2 x_{\mu} x^{\nu} \partial_{\nu} 
\Phi + 2 d x_{\mu} \Phi - x_{\nu} x^{\nu} \partial_{\mu} \Phi - 2 
\mathrm{i} x^{\nu} S_{\mu \nu} \Phi) .   \label{tgfsct}
\end{equation}
The transformation of the metric (\ref{sctvame}) can be written in 
this form, provided that we use the appropriate representation of the 
spin generator for second rank tensors~\cite{ramond}. Substituting 
(\ref{tgfsct}) and (\ref{sct}) in (\ref{newtvl}), we obtain
\begin{eqnarray}
\Delta_S {\mathcal L} &=& 2 c^{\nu} x_{\nu} \bigg(4 
{\mathcal L} - d \frac{\partial {\mathcal L}}{\partial \Phi} \Phi - 
(d + 1) \frac{\partial {\mathcal L}}{\partial (\partial_{\mu} \Phi)} 
\partial_{\mu} \Phi \nonumber \\ &+& {\mathcal T}^{\mu}{}_{\mu} \bigg) 
+ 2 c^{\nu} 
\frac{\partial {\mathcal L}}{\partial (\partial_{\mu} \Phi)} 
(\mathrm{i} S_{\nu \mu} \Phi - d \eta_{\mu \nu} \Phi) \; ,
\label{newtvlsct}
\end{eqnarray}
where we used (\ref{sctvame}), (\ref{vlsemt}) and that the Lagrangian 
is invariant under translations (\ref{pdl}) and Lorentz 
transformations~\cite{cacoja,coja,blago}
\begin{eqnarray}
\mathrm{i} \frac{\partial {\mathcal L}}{\partial \Phi} S_{\mu \nu} 
\Phi &+& \mathrm{i} 
\frac{\partial {\mathcal L}}{\partial (\partial_{\rho} \Phi)} 
\partial_{\rho} (S_{\mu \nu} \Phi) \nonumber \\ &-& 
\frac{\partial {\mathcal L}}{\partial (\partial_{\rho} \Phi)} 
(\eta_{\mu \rho} \partial_{\nu} \Phi - \eta_{\nu \rho} \partial_{\mu} 
\Phi) = 0 \; .   \label{pdl2}
\end{eqnarray}
Therefore, in order to have invariance under the special conformal 
transformations, $\Delta_S {\mathcal L} = 0$, we must have, 
first of all, invariance under dilatations transformations, 
(\ref{newpdl3}). Then, the remaining condition is just
\begin{equation}
{\mathcal S}^{\mu}{}_{\nu \mu} - d 
\frac{\partial {\mathcal L}}{\partial (\partial_{\mu} \Phi)} 
\eta_{\mu \nu} \Phi = 0 ,   \label{tracest}
\end{equation}
where we used the definition of the {\it Spin} tensor (\ref{spintgf}). 
Consequently, if we can write the ``trace'' of the {\it Spin} tensor 
in the above manner, then, the action will be invariant under the 
special conformal transformations.

Now, substituting (\ref{sct}) and (\ref{tgfsct}) in (\ref{ncc}), and 
using (\ref{sctvame}) and (\ref{vlsemt}), we obtain that
\begin{equation}
\partial_{\mu} K^{\mu}{}_{\nu} = 2 x_{\nu} {\mathcal T}^{\mu}{}_{\mu} ,
\label{cosctcu}
\end{equation}
where
\begin{equation}
K^{\mu}{}_{\nu} = 2 x_{\nu} x^{\rho} t^{\mu}{}_{\rho} + 2 d x_{\nu} 
\frac{\partial {\mathcal L}}{\partial (\partial_{\mu} \Phi)} \Phi - 
x_{\rho} x^{\rho} t^{\mu}{}_{\nu} - 2 x^{\rho} 
{\mathcal S}^{\mu}{}_{\nu \rho} ,   \label{sccu}
\end{equation}
is the {\it special conformal current}. Consequently, the special 
conformal current is not conserved unless the trace of the {\it SEMT} 
vanishes.

In~\cite{cacoja}, they imposed (\ref{tracest}) to be a divergence, and 
constructed a new improved energy-momentum tensor. However, if we 
impose (\ref{tracest}) to vanish, the improved energy-momentum tensor 
becomes, essentially, the {\it SEMT} (\ref{semt}). This is consistent 
with (\ref{codicu}) and (\ref{cosctcu}), where instead of the improved 
energy-momentum tensor, there appears the {\it SEMT}. In fact, if we 
substitute (\ref{semt}) in (\ref{dilacu}) and (\ref{sccu}), and use 
(\ref{tracest}), after a partial integration we obtain that
\begin{equation}
D^{\mu} = x^{\nu} {\mathcal T}^{\mu}{}_{\nu} + \partial_{\rho} 
\left(\frac{1}{2} x^{\nu} \varphi^{\rho \mu}{}_{\nu} \right) ,
\label{dilacu2}
\end{equation}
and
\begin{eqnarray}
K^{\mu}{}_{\nu} &=& (2 x_{\nu} x^{\rho} - x_{\sigma} x^{\sigma} 
\delta^{\rho}{}_{\nu}) {\mathcal T}^{\mu}{}_{\rho} \nonumber \\ &+& 
\partial_{\sigma} \left[\left(x_{\nu} x^{\rho} - \frac{1}{2} x_{\lambda} 
x^{\lambda} \delta^{\rho}{}_{\nu} \right) \varphi^{\sigma \mu}{}_{\rho} 
\right] . \label{sccu2}
\end{eqnarray}
As the lasts terms in (\ref{dilacu2}) and (\ref{sccu2}) are just the 
divergence of an anti-symmetric tensor, we can discard them because 
they do not affect the ``conservation'' of the currents. It is worth 
to note that we could also put the conformal currents in a simple manner, 
similar to that in~\cite{cacoja}, without the necessity of constructing 
an improved energy-momentum tensor. Consequently, it seems that it is the 
{\it SEMT} that plays a fundamental role in the conformal transformations.

\section{Conformal Transformations of the Fields}

Now, we will apply the above considerations to the massive scalar and 
spinor field, and to the electromagnetic field. Let us begin, then, 
with the massive scalar field. The Lagrangian of the scalar field is
\begin{equation}
{\mathcal L} = \frac{1}{2} \eta^{\mu \nu} \partial_{\mu} \phi 
\partial_{\nu} \phi - \frac{1}{2} m^2 \phi^2 .   \label{scala}
\end{equation}
As for the scalar field ${\mathcal T}^{\mu}{}_{\mu} = - 2 {\mathcal L} 
+ m^2 \phi^2$, then,  substituting (\ref{scala}) in (\ref{newtvld}) we 
obtain
\begin{equation}
\Delta_D {\mathcal L} = - 2 a d {\mathcal L} .
\end{equation}
We see that the mass term does not appear in the variation of the 
action. This is because the usual mass term that would appear from the 
variation of the field $\phi$ is {\it canceled} by the mass term that 
comes from the {\it SEMT}, which is the variation with respect to the 
metric. Invariance of the scalar field action under dilatations, 
$\Delta_D {\mathcal L} = 0$, implies that
\begin{equation}
d (\phi) = 0 .
\end{equation}
Note that as the scale dimension is not related with the natural 
dimension, the fact that the scale dimension of the scalar field $\phi$ 
vanishes does not means that its natural dimension also vanishes. 
From (\ref{fvpd}) we see that the change of the scalar field under 
dilatations becomes
\begin{equation}
\bar{\delta}_D \phi = - a x^{\mu} \partial_{\mu} \phi = - \delta_D 
x^{\mu} \partial_{\mu} \phi ,
\end{equation}
where we used (\ref{dila}). The last term is the usual transformation 
law of the scalar field under general coordinate transformations. To 
prove invariance under the special conformal transformations, we just 
need to show that (\ref{tracest}) is valid. As $S_{\mu \nu} \phi = 0$, 
then, the {\it Spin} tensor vanishes. But, as $d (\phi) = 0$, the 
second term in (\ref{tracest}) vanishes too. Therefore, the massive 
scalar field is also invariant under the special conformal 
transformations. (This can be seen, too, by explicitly substituting 
(\ref{scala}) in (\ref{newtvlsct})). The transformation of the scalar 
field under special conformal transformations is, then,
\begin{equation}
\bar{\delta}_S \phi = - c^{\nu} (2 x_{\nu} x^{\mu} \partial_{\mu} 
\phi - x_{\mu} x^{\mu} \partial_{\nu} \phi) = - \delta_S x^{\mu} 
\partial_{\mu} \phi ,
\end{equation}
where we used (\ref{sct}). Therefore, the transformation law of the 
scalar field under the conformal transformations can be obtained from 
the transformation law of the scalar field under general coordinate 
transformations.

The spinor field Lagrangian is
\begin{equation}
{\mathcal L} = \frac{\mathrm{i}}{2} \left(\bar{\psi} \gamma^{\mu} 
\partial_{\mu} \psi - \partial_{\mu} \bar{\psi} \gamma^{\mu} \psi 
\right) - m \bar{\psi} \psi .   \label{spila}
\end{equation}
As for the spinor field ${\mathcal T}^{\mu}{}_{\mu} = - 3 {\mathcal L} + 
m \bar{\psi} \psi$, then, substituting (\ref{spila}) in (\ref{newtvld}) 
we obtain
\begin{equation}
\Delta_D {\mathcal L} = - 2 a d {\mathcal L} .
\end{equation}
We also see that the mass term of the {\it SEMT} canceled the usual mass 
term. Invariance under dilatations, $\Delta_D {\mathcal L} = 0$, requires 
that
\begin{equation}
d (\psi) = 0 .
\end{equation}
Again, while the scale dimension of the spinor field $\psi$ vanishes, its 
natural dimension does not. Similar to the scalar field, the 
transformation of spinor field under dilatations is
\begin{equation}
\bar{\delta}_D \psi = - a x^{\mu} \partial_{\mu} \psi = - \delta_D 
x^{\mu} \partial_{\mu} \psi ,
\end{equation}
since the spinor field transform as a scalar under general coordinate 
transformations too. To see the invariance under the special conformal 
transformations, we just need to show that (\ref{tracest}) is 
satisfied. As the {\it Spin} tensor of the spinor field is totally 
anti-symmetric in its three indices, its trace vanishes. But, as 
$d (\psi) = 0$, both terms in (\ref{tracest}) vanishes. Therefore, the 
spinor field is also invariant under the special conformal 
transformations.

From (\ref{dilacu}) and (\ref{codicu}), we can see that for $d = 0$,
\begin{equation}
\partial_{\mu} D^{\mu} = t^{\mu}{}_{\mu} = {\mathcal T}^{\mu}{}_{\mu}.
\label{codid0}
\end{equation}
This is clearly true for the scalar field, because for this field the 
{\it CEMT} coincides with the {\it SEMT}. For the spinor field, the 
{\it SEMT} is just the symmetrized version of the {\it CEMT}, unless 
of a $1/2$ factor, so that (\ref{codid0}) is also true.

For the electromagnetic field $A_{\mu}$, all the usual considerations 
are still valid, because for this field 
${\mathcal T}^{\mu}{}_{\mu} = 0$. It is not difficult to show from 
(\ref{newtvld}) that the condition for invariance under dilatations 
requires that $d (A_{\mu}) = 1$. The scale dimension and the natural 
dimension of the electromagnetic field $A_{\mu}$ just coincide. 
We can show, too, that the trace of 
the {\it Spin} tensor of the electromagnetic field can be written as 
in (\ref{tracest}), so that, the electromagnetic field is invariant 
under the special conformal transformations. As with the scalar field, 
we can obtain the transformation law of the electromagnetic field 
$A_{\mu}$ under dilatations (\ref{fvpd}) and special conformal 
transformations (\ref{tgfsct}) from the transformation law of vectors 
fields under general coordinate transformation,
\begin{equation}
\bar{\delta} A_{\mu} = - \delta x^{\nu} \partial_{\nu} A_{\mu} - 
A_{\nu} \partial_{\mu} \delta x^{\nu} ,
\end{equation}
just substituting (\ref{dila}) and (\ref{sct}), respectively. It is 
interesting to note that as the metric is not invariant under 
dilatations, for example, we have that
\begin{equation}
\bar{\delta}_D A^{\mu} = \bar{\delta}_D (\eta^{\mu \nu} A_{\nu}) = - a 
x^{\nu} \partial_{\nu} A^{\mu} + a A^{\mu} ,
\end{equation}
where we used that $d (\eta^{\mu \nu}) = - 2$. This can be put in the 
form of (\ref{fvpd}) if $d (A^{\mu}) = - 1$. 

\section{Final Remarks}

According to the conformal Killing equation (\ref{cke}), we are 
looking for coordinate transformations of the metric, the 
{\it l.h.s.} of (\ref{cke}), that changes the form of the metric 
according to (\ref{contrame}), the {\it r.h.s.} of (\ref{cke}). 
Therefore, the conformal transformations are not isometries of the 
metric, because they do not obey a Killing equation. Hence, not only 
any general field $\Phi$, but also spacetime changes under the conformal 
transformations. In this sense, there seems to be no reason to consider 
also a Weyl rescaling of the metric to impose that the metric does not 
varies, as is usually done~\cite{wess2,blago}. Not only because the 
metric changes indeed, but also because it is considered only a Weyl 
rescaling of the metric, and not of the other fields.

When we take the variation of the metric in the variation of the 
action, we acquire an extra term which is, basically, the {\it SEMT}. 
Then, through a similar procedure to the usual Noether theorem, we see 
that we no longer obtain a conserved current, as can be seen from 
(\ref{ncc}). When the symmetry transformation is an isometry of the 
metric, we do obtain conserved currents. For the case of the conformal 
transformations, only fields for which the trace of the {\it SEMT} 
vanishes have conserved currents.

Similar results on the ``conservation'' of conformal currents were 
obtained in~\cite{cacoja} through the definition of a new improved 
energy-momentum tensor. Nevertheless, provided that (\ref{tracest}) 
vanishes, instead of being a divergence, this improved energy-momentum 
tensor becomes essentially the {\it SEMT}. Then, there would be no 
need to modify General Relativity.

The variation of the metric introduces a mass term, through the trace 
of the {\it SEMT}, which just cancels the mass term that comes from 
the variation of the fields. Consequently, even massive fields can be 
invariant under the conformal transformations, but, we should  stress 
one more time that this does not means that they have a conserved 
current. Then, the massive scalar field does not play any fundamental 
role, as was previously thought. While classically dilatation can now 
be a symmetry of the fields, quantically it can not~\cite{wess}. As 
$d (\eta^{\mu \nu}) = - 2$, now we have
\begin{equation}
[D, P^2] = [D, \eta^{\mu \nu} P_{\mu} P_{\nu}] = 4 \mathrm{i} P^2 .
\end{equation}
We still obtain that the mass spectrum is continuous, or that all 
masses vanishes. 

The invariance condition under dilatations (\ref{newtvld}) requires 
that $d = 0$ for fields that transforms as scalars under general 
coordinate transformations, and $d = 1$ for fields that transforms as 
vectors under general coordinate transformations. As 
$d (\eta_{\mu \nu}) = 2$, and $\eta_{\mu \nu}$ is a second rank 
tensor, it seems that the scale dimension $d$ is related to the tensorial 
character of the fields. We should remark that, as pointed 
in~\cite{carru,coja}, the scale dimension $d$ is not related to the 
natural dimensions of the fields. They just used to coincide. Now, that 
we are taking the variation of the metric, we see that they do not 
coincide anymore. 
We should note that the Lagrangian 
${\mathcal L}$ is, in fact, a scalar density, 
${\mathcal L} = \sqrt{- \eta} L$. Then, from (\ref{divame}) we can 
show that $\bar{\delta}_D \sqrt{- \eta} = - 4 a \sqrt{- \eta}$, so 
that from (\ref{newfvl3}) we see that
\begin{equation}
\bar{\delta}_D L = - a x^{\mu} \partial_{\mu} L = - \delta_D x^{\mu} 
\partial_{\mu} L .
\end{equation}
So, $L$ transforms as a scalar.

Finally, we have seen that the transformation 
law of the fields under the conformal transformations can be obtained 
from the transformation law of the fields under general coordinate 
transformations. This is as it should be, because the conformal 
transformations are, in fact, coordinate transformations. The fact that 
the action of the fields turns out now to be invariant under these 
transformations should also be expected, because the action should 
be invariant under coordinate transformations.

\begin{acknowledgments}
I would like to thanks Profs. J. G. Pereira and J. A. Helayel-Neto 
for very useful discussions, and B. C. Vallilo for some earlier 
comments on conformal transformations. I also thank FAPESP-Brazil 
for financial support.
\end{acknowledgments}

\end{document}